\documentstyle[,psfig,floats,aps,prb,twocolumn]{revtex}
 
\begin{document}

\wideabs{  
\title{Level Set Approach to Reversible Epitaxial Growth}

\author{M. Petersen$^{1,2}$, C. Ratsch$^2$, R. E. Caflisch$^2$, 
        and A. Zangwill$^1$}

\address{
$^1$School of Physics, Georgia Institute of Technology, Atlanta, GA 30332\\
$^2$Department of Mathematics, UCLA, Los Angeles, CA 90095-1555\\
}
\date{\today}
\maketitle
\begin{abstract}
  We generalize the level set approach to model epitaxial growth to
  include thermal detachment of atoms from island edges. This means
  that islands do not always grow and island dissociation can
  occur. We make no assumptions about a critical nucleus. Excellent
  quantitative agreement is obtained with kinetic Monte Carlo
  simulations for island densities and island size distributions in
  the submonolayer regime.
\end{abstract}

\pacs{PACS numbers:}
}
\section{Introduction}

At the present time, there is no practical approach to epitaxial
growth modelling that bridges the gap between microscopic and
macroscopic length scales.  Rate equations offer some hope, \cite{RE}
but their reliance on uncontrolled mean-field approximations remains a
serious obstacle.  Kinetic Monte Carlo (KMC) simulations are very
popular, \cite{KMC} but scale-up to the micron range is very doubtful,
even with future supercomputers.

One approach to the multiple-scale problem uses an atomic description
in the vertical (growth) direction and a continuum description in the
lateral directions. Specifically, the random walk of individual atoms
on a flat terrace is replaced by the solution of a diffusion equation
for the monomer density on each terrace. This is not a new idea
\cite{BCF}, but its recent rebirth in the context of the {\it level
set} (LVST) method \cite{CaGy99,ChKa00,Cho00} is particularly
promising in light of the relatively low computational cost needed to
treat arbitrarily complicated surface morphologies. So far, good
success has been achieved for {\it irreversible} epitaxial growth
where LVST calculations quantitatively reproduce the results of KMC
calculations for the distribution of two-dimensional islands in the
submonolayer regime. \cite{RaGy00}

The purpose of this paper is to extend the LVST method to the case of
{\it reversible} epitaxial growth where thermal detachment of atoms
from island edges is allowed. This step is necessary if one hopes to
produce a model that is relevant to growth at elevated
temperatures. Moreover, a reversible LVST growth model has significant
computational advantages over a reversible KMC model. This is so
because KMC keeps track of every detaching atom, including those that
eventually return to the island from whence they came.  Such events
leave the system unchanged overall \cite{caveat} and slow down the
simulation significantly. By contrast, the reversible LVST scheme we
develop below replaces these events by their time average and so
includes only those detachments that do not lead to subsequent
reattachments.  Moreover, because of the mean field approach, a large
number of detachment events can be treated within a single simulation
time step.

\section{Method}

\subsection{Level Sets}

The level set method \cite{OsSe88} models the time evolution of
arbitrarily shaped objects in $n$ dimensions that can undergo
topological changes. In this paper, the relevant objects are
two-dimensional islands and topological changes occur due to
nucleation, dissociation, and coalescence.  The key idea is to
represent a curve or interface $\Gamma$ in $\Re^n$ by the level $k$ of
a function $\phi({\rm\bf x},t)$
\begin{eqnarray}
\Gamma_k & = & \{ {\rm\bf x} \, : \, \phi({\rm\bf x},t) = k \}  
\qquad {\rm\bf x} \in \Re^n \quad .
\label{lvst_1}
\end{eqnarray} 
Here, $\Gamma_k$ is the set of closed curves that constitute the
perimeters of the islands with height $(k+1)a$ ($a$ is the lattice
parameter).

\begin{figure}[ht]
\vspace{-.8cm}
\begin{center}
\psfig{figure=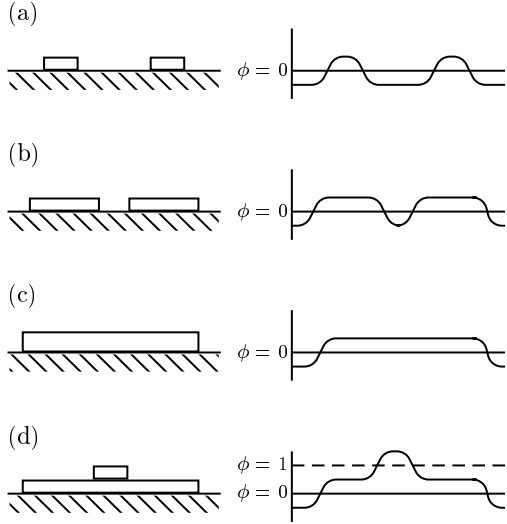,width=7cm}
\end{center}
\caption{\label{scem_LS}
   Schematic illustration on mapping island configurations during
   growth (left panel) onto a LS function $\phi$.}
\end{figure}

Fig.~\ref{scem_LS} illustrates the level set description of a typical
epitaxial growth scenario. The left panel is a side view of two
islands on a terrace (a) that grow to a precoalesence state (b) and
subsequently merge (c). Later, a new island nucleates on top (d).  The
right panel shows the corresponding level set functions $\phi$. Note
that it is not $\phi$ that represents the surface morphology, but only
the level sets ($\phi$ = 0 and $\phi$ = 1).
%
% Fig 1
%

The motion of $\Gamma$ is partly deterministic and partly stochastic.
There is a deterministic part because a mean field theory 
is sufficient to model the time-average of many of the physical
processes that contribute to growth. These effects are built into a
velocity function
$v_n ({\rm\bf x},t)$ that evolves the function $\phi({\rm\bf x},t)$
in time according to the
partial differential equation
\begin{eqnarray}
\dot{\phi}({\rm\bf x},t) + v_n ({\rm\bf x},t) 
|\nabla\phi({\rm\bf x},t)| & = & 0. \quad
\label{lvst_3}
\end{eqnarray}
As the notation suggests, $v_n$ is the component of the growth
velocity in the direction of the local surface normal 
${\rm\bf n}=\nabla\phi/|\nabla\phi|$.  
The stochastic motion of $\Gamma$ is
associated with nucleation events and small-island dynamics. There is
no unique algorithm to incorporate these effects into
$\dot{\phi}({\rm\bf x},t)$. The particular choice we make is explained
in detail below.

\subsection{Deterministic Evolution}

It is convenient to write the velocity $v_n$ in Eq.~(\ref{lvst_3})
in the form
\begin{eqnarray}
v_n  & = & v^{\rm att} - v^{\rm det}
\quad 
\label{v_rev}
\end{eqnarray}
where $v^{\rm att}$ accounts for attachment processes that grow
islands and $v^{\rm det}$ accounts for detachment processes that
shrink islands.  The first of these is proportional to the diffusive
flux of atoms that approach an island edge from its bounding
terraces. Therefore, if $D$ is the surface diffusion constant and
$\rho({\rm\bf x},t)$ is the adatom density, mass conservation gives
\begin{eqnarray}
v^{\rm att} =  a^2\, D\left( 
\frac{\partial \rho}{\partial n}|_{\rm \mbox{terrace}} -
\frac{\partial \rho}{\partial n}|_{\rm \mbox{top of island}} 
\right). \quad \label{v_grow}
\end{eqnarray}
We compute the required density from the mean-field, driven, diffusion
equation 
\begin{eqnarray}
\dot{\rho}({\rm\bf x},t) & = & D\nabla^2 \rho({\rm\bf x},t) 
+ F - 2\frac{dN_{\rm nuc}}{dt}. \quad 
\label{ad_solv}
\end{eqnarray}

The loss term in Eq.~(\ref{ad_solv}),
\begin{eqnarray}
\frac{dN_{\rm nuc}(t)}{dt} & = & D \sigma_1 \int_{\Omega} \rho({\rm\bf x},t)^2 d^2x,
\label{nucl}
\end{eqnarray}
accounts for the dimers that nucleate as a result of binary collision
between monomers.  The total simulation area is $\Omega$ and
$\sigma_1$ is the so-called "capture number" for an
adatom.\cite{BaCh94} We solve Eq.~(\ref{ad_solv}) subject to the
boundary condition that $\rho=0$ at every point on $\Gamma$. This
differs from the boundary condition usually used for reversible
aggregation \cite{BaZa97} because we have elected to incorporate all
detachment effects into the velocity $v^{\rm det}$ (see Appendix A).

To find an explicit expression for $v^{\rm det}$, we note first that
most particles that detach from an island are driven back to that
island by the diffusion field. \cite{BaZa97} Our interest here is
those particles that detach without subsequent reattachment, {\it
i.e.}, those that escape from the "capture zone" of the island. This
is so because, by definition, the adatoms in the capture zone of a
given island are guaranteed to attach to that island eventually.  A
relevant quantity is thus $p_{\rm esc}$, the probability that a
detached particle reaches the border of the capture zone.  The
positions of these borders are easily calculated (the locus of points
where $\nabla \rho =0$) and so is $p_{\rm esc}$. We find (Appendix B)
that
\begin{eqnarray}
p_{\rm esc} & = & \frac{\log((R_{\rm is} + 
a)/R_{\rm is})}{\log(R_{\rm cz}/R_{\rm is})}
\label{p_esc}
\end{eqnarray}
where $R_{\rm is}$ and $R_{\rm cz}$ are the radii of the circularly
averaged island and capture zone.

We now define an effective escape rate per unit length of
island perimeter as 
\begin{eqnarray}
R_{\rm det} &  = & D_{\rm det}\,p_{\rm esc}\,\lambda
\quad 
\label{int_1}
\end{eqnarray}
where $D_{\rm det}$ is an effective detachment rate and $\lambda$ is
the linear density of detaching particles (singly coordinated edge
atoms).  We use $\lambda=2$ for ``small'' islands and the expression
\cite{CaE99}
\begin{eqnarray}
\lambda & = & \frac{16}{3}\,
\left(\frac{16}{15}\,v^{\rm att}\,
\frac{1}{D_{\rm edg}}\right)^{\frac{2}{3}}
\quad 
\label{kink}
\end{eqnarray}
for ``large'' islands. The distinction between ``large'' and ``small''
islands will be made clear below. $D_{\rm edg}$ is the edge diffusion
constant.\cite{comment-adatom} From $R_{\rm det}$ we get the desired
expression for the detachment velocity that enters Eq.~(\ref{v_rev}):
\begin{eqnarray}
{\rm\bf v}^{\rm det} 
& = & a^2 R_{\rm det} 
  =   a^2 D_{\rm det}\,p_{\rm esc}\,\lambda \quad 
\label{vs_def}
\end{eqnarray}
Significantly (see below), there is negligible extra computational
overhead needed to incorporate detachment in this way.

It remains only to find a home for the atoms that escape from the
islands.  In the spirit of the mean-field approximation, we return
them to the adatom pool by simply augmenting the external flux. That
is, the variable $F$ in Eq.~(\ref{ad_solv}) is
\begin{eqnarray}
F & = & F_0 + F_{\rm rev} \quad 
\label{flux_rev}
\end{eqnarray}
where $F_0$ is the deposition flux and
\begin{eqnarray}  
F_{\rm rev} & = & \frac{1}{\Omega}\int_{\Gamma} R_{\rm det} d\Gamma \quad
\label{loss_total}
\end{eqnarray}  
is the escape rate of atoms from all island edges. In this integral, 
$\Gamma$ runs over all level sets of $\phi({\rm\bf x},t)$ (see
Eq.~(\ref{lvst_1})).

\subsection{Stochastic Evolution}

A nucleation event occurs when the variable $N_{\rm nuc}$ in
Eq.~(\ref{nucl}) becomes larger than the next integer. This implies
that $\phi({\rm\bf x},t)$ increases by a discrete amount at a discrete
point. In the interest of numerical stabilty, we smooth out this
increase over several points on the numerical gid used to solve
Eq.~(\ref{ad_solv}). The exact position where the dimer nucleates is
chosen randomly with the integrand of Eq.~(\ref{nucl}) as a weight
factor.\cite{RaGy00}

Randomness is also important for detachment from ``small'' islands
that consist of only a few atoms.  Our approach is to choose an island
area $A_{\rm cut}$ and treat all islands smaller than this size
statistically.  Thus, in a given time interval, we use
Eq.~(\ref{int_1}) to calculate the total number of adatoms that detach
from all islands smaller than $A_{\rm cut}$. This corresponds to a
total area-loss $A_{\rm loss}$. If $A_{\rm loss}>A_1$, the area
occupied by one atom, we detach an atom from one of the islands
smaller than $A_{\rm cut}$. The specific island that loses an atom is
chosen randomly with $p_{\rm esc}$ as a weight factor. We then
decrement $A_{\rm loss}$ by $A_1$ and repeat the process until $A_{\rm
loss} < A_1$. This value is stored and added to the loss that occurs
in the next time interval. If a detachment process leads to an island
smaller than a dimer, we dissociate the dimer and decrement $A_{\rm
loss}$ accordingly.

We emphasize that $A_{\rm cut}$ is {\it not} the area of a critical
nucleus, {\it i.e.}, an island that is absolutely stable against
breakup. Instead, $A_{\rm cut}$ is merely the size of the smallest
``large'' island which we use as a parameter to switch between
statistical and continuous detachment.  In the context of our
approach, the critical nucleus is defined by the condition $v^{\rm
att}=v^{\rm det}$ (see Eq.~(\ref{v_rev})).

\section{Results}

\subsection{Parameter Choices \& Systematics}

\begin{figure}[th]
\begin{center}
\psfig{figure=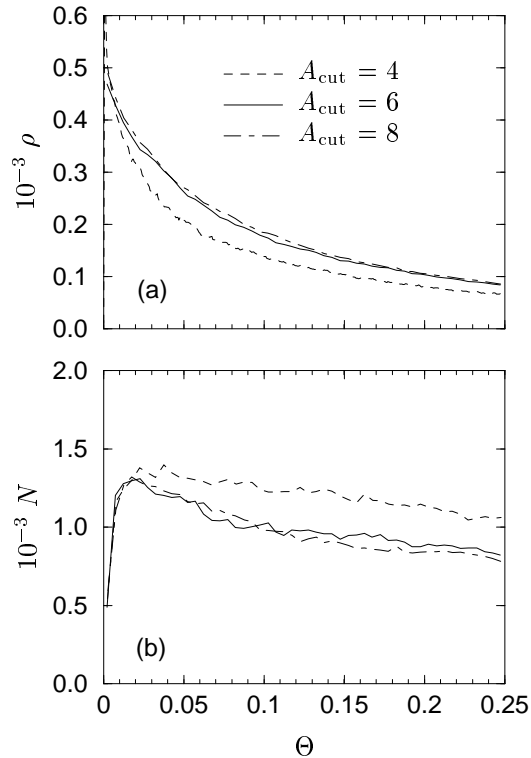,width=7cm}
\end{center}
\caption{\label{fig_acut} 
   Influence of $A_{\rm cut}$ on the adatom density $\rho$ (a) and the island density
   $N$ (b) as a function of coverage $\Theta$ for $D_{\rm det}$/$D$ =
   0.001. Data have been averaged over 10 runs.}
\end{figure}

All the LVST simulations reported here use the value
$D/F_0=10^6$. $D_{\rm edg}$ in Eq.~(\ref{kink}) was chosen as 10$^4$
and calculations where carried out on a 200$\times$200 lattice
represented on a numerical grid of 568$\times$568 points. To determine
$A_{\rm cut}$, we compared runs with different choices for this
parameter and looked for stabilization of the physical results. Thus,
Fig~\ref{fig_acut} compares island and adatom densities as a function
of coverage $\Theta$ obtained for $A_{\rm cut}$ = 4,6, and 8 at a
detachment rate of $D_{\rm det}$/$D$ = 0.001.  Based on this data and
related statistical tests, we find that our results are independent of
$A_{\rm cut}$ if $A_{\rm cut}\ge 6$. Therefore we use $A_{\rm cut}=6$
in all subsequent simulations.
%
% Fig 2
%

\begin{figure}[th]
\begin{center}
\psfig{figure=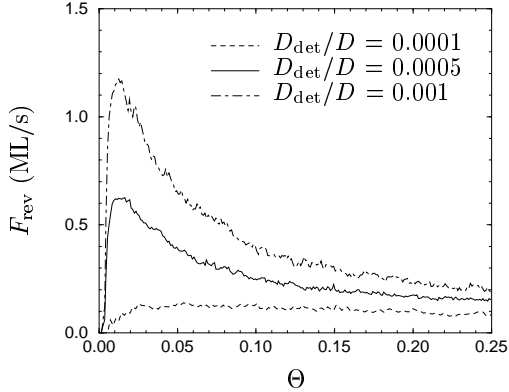,width=7cm}
\end{center}
\caption{\label{fig_rtot}
  Escape rate of atoms from all island edges $F_{\rm rev}$ (see
  Eq.~(\ref{loss_total})) as a function of coverage $\Theta$ for
  different values of $D_{\rm det}$/$D$. Simulation parameters as in
  Fig.~\ref{fig_acut}, except $A_{\rm cut}=6$.}
\end{figure}

Fig.~\ref{fig_rtot} illustrates the role of detachment during growth.
It shows $F_{\rm rev}$ from Eq.~(\ref{flux_rev}) as a function of
$F_{0}t$ for different values of $D_{\rm det}$/$D$.  The fact that
$F_{\rm rev}$ has its highest values at the earliest stages of growth
(for the higher values of $D_{\rm det}$/$D$) shows that most of the
effective detachment takes place when the islands are quite small.
Indeed, as islands grow bigger, fewer particles escape from the island
boundaries since the number of escaping particles relative to the
perimeter of the growing islands becomes smaller and smaller.
%
% Fig 3
%

\subsection{Comparison to KMC}

As a critical test, we compared our LVST results directly with KMC
simulations.\cite{KMC,ross00}
%
% ------- old -----
%
% The parameters for the latter were chosen to give a migration rate of
% $D=10^6$ for isolated adatoms and 
% $D_{\rm det}$ as indicated below for singly-coordinated atoms.
%
% ------ new ------
%
In our KMC simulations adatoms are allowd to hop to a nearest-neighbor
site at a rate $r_n = D \exp(-nE_N/k_{\rm B}T)$, where $n$ is the
number of nearest-neighbours, $k_{\rm B}$ the Boltzmann-constant, and
$T$ the temperature. Adatoms are deposited at a rate $F_0$. In all
simulations presented in this paper the ratio $D$/$F_0$ is set to
$10^6$.  The energy barrier $E_N$ is chosen such that $D_{\rm det} =
D\exp(-E_N/k_{\rm B}T)$. In addition, singly coordinated edge-atoms
are allowed to diffuse along the step edge at a rate $D_{\rm edg}$.
We chose $D_{\rm edg}$=$D_{\rm det}$, but the results were not
sensitive to this parameter as long as the islands were compact (as
our LVST model assumes).

\begin{figure}[hbt]
  \begin{center}
  \psfig{figure=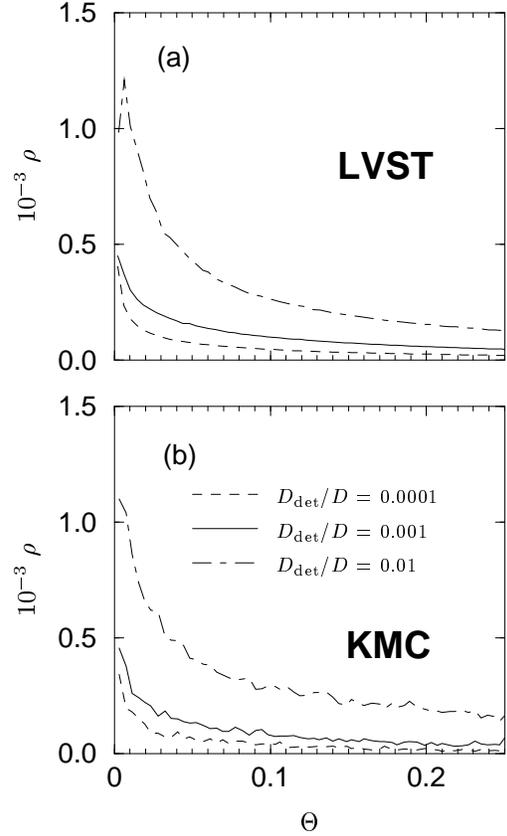,width=7cm}
  \end{center}
  \caption{\label{compare_rho}
    Adatom densities as obtained by the level set method (a) and KMC
    (b). In the LVST calculation all parameters are as in
    Fig.~\ref{fig_acut}, except $A_{\rm cut}=6$.}
\end{figure}

\begin{figure}[h]
  \begin{center}
  \psfig{figure=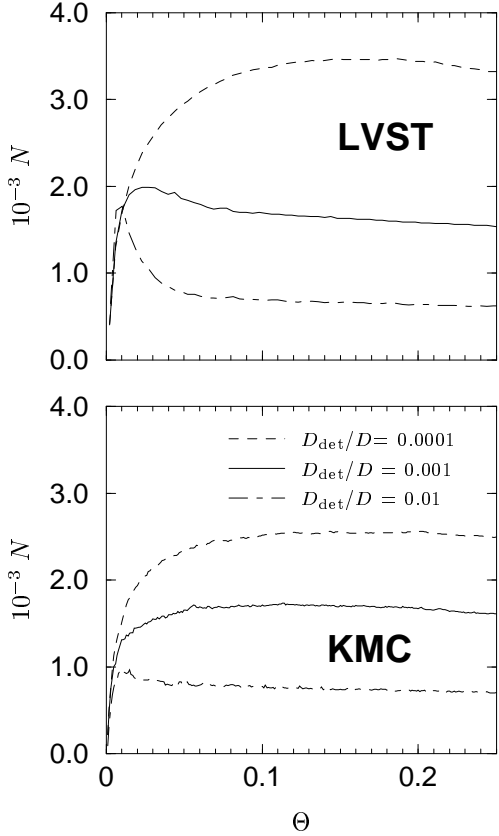,width=7cm}
  \end{center}
  \caption{\label{compare_N}
    Island densities $N$ as obtained by the level set method
    (a) and KMC (b). Parameters as in Fig.~\ref{compare_rho}.
}
\end{figure}

\begin{figure}[th]
   \psfig{figure=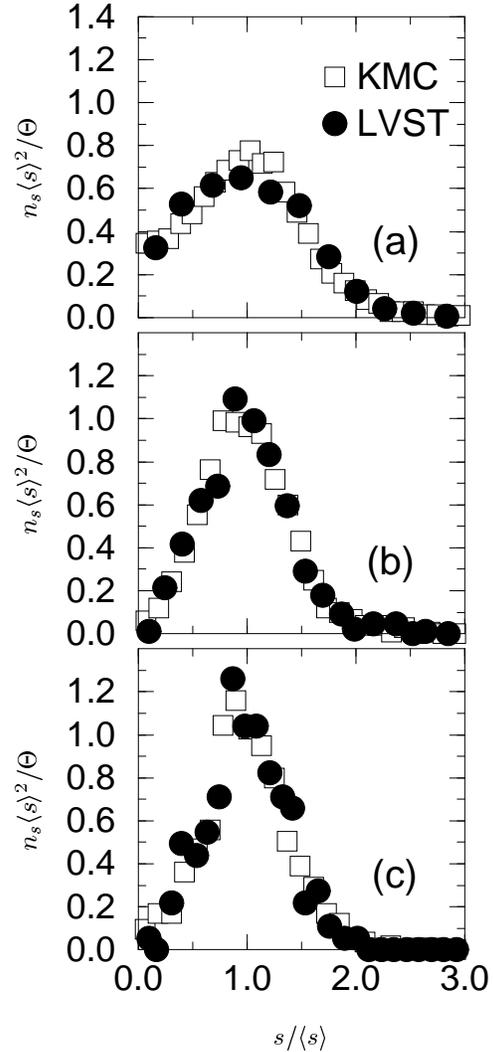,width=7cm}
  \caption{\label{CSD}
     Island size distributions where $n_s$ is the density of islands
     of size $s$, $\langle s\rangle$ is the average island size, and
     $\Theta$ is the coverage. Closed circles: Level set result, open
     squares: KMC. Detachment rates are in Fig. (a): $D_{\rm
     det}$/$D$=0.0001, in Fig. (b): $D_{\rm det}$/$D$=0.0005, and in
     Fig. (c): $D_{\rm det}$/$D$=0.001.  Data have been sampled at
     $\Theta$ = 0.25. Other parameters as in Fig.~\ref{compare_rho}.
     }
\end{figure}

Fig.~\ref{compare_rho}, Fig.~\ref{compare_N}, and Fig.~\ref{CSD}
respectively show the adatom density, island density, and island size
distributon obtained by both methods as a function of detachment
rate. These curves agree with previous reversible KMC simulations
\cite{BaZa97,RZSV,RaSm95} that discuss, {\it e.g.}, the physical
origin of the observed saturation of the island densities and the
sharpening of the island size distributions.  Evidently, there is
semi-quantitative agreement between LVST and KMC. The size
distributions results are particularly notable because they reflect
information about spatial correlations that are averaged over to get
$\rho$ and $N$.
%
% Fig 4
%

%
% Fig 5
%

The only disagreement we find between LVST and KMC is for the
saturation value of the island density for the smallest value of the
detachment rate. We understand this based on recent research
\cite{RaKa01} with irreversible growth where a corresponding
disagreement {\it vanishes} when the $\rho=0$ boundary condition for
Eq.~(\ref{ad_solv}) is applied not at the true perimeter of each
island (as we do) but instead at a closed boundary that exceeds the
perimeter everywhere by one lattice constant. The disagreement
disappears altogether when the island density is small. This is
consistent with our observations because we get good agreement with
KMC when the detachment rate is large (and the island density is
small).

\begin{figure}[h]
  \begin{center}
   \psfig{figure=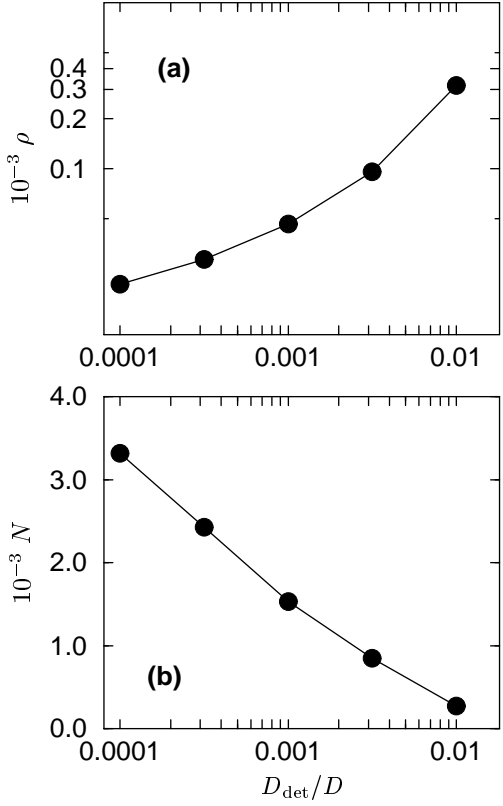,width=7cm}
  \end{center}
  \caption{\label{equi_fig}
  Equilibrium adatom density $\rho$ (a) and equilibrium density of islands $N$
  (b).  Data have been sampled at 0.25 coverage. Other parameters as
  in Fig.~\ref{compare_rho}.
}
\end{figure}

Fig.~\ref{equi_fig} shows some systematic features of reversible
growth as a function of the effective detachment rate. The LVST data
(collected at 0.25 ML coverage) show that the average adatom density
increases linearly with $D_{\rm det}$ while the average island density
decreases exponentially with $D_{\rm det}$.  Unfortunately, we have
been unable to derive theses interesting results analytically using
rate equations.

\begin{figure}[h]
  \begin{center}
   \psfig{figure=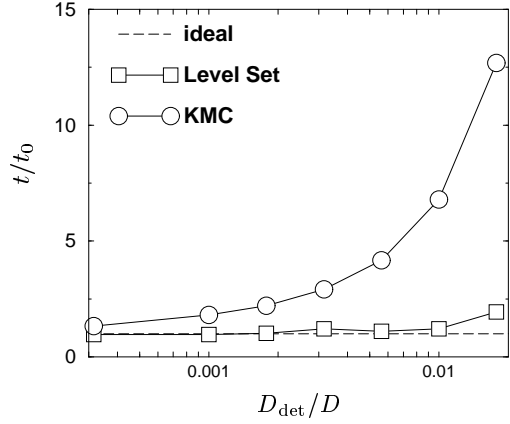,width=7cm}	
  \end{center}
  \caption{\label{scaling}
    Scaling of the level set and KMC method as a function of
    detachment rate. Runtimes $t$ are normalized to the runtime of the
    irreversible case $t_0$.  Parameters as in Fig.~\ref{compare_rho}.  
} 
\end{figure}

The data in Fig.~\ref{CSD} was easy to obtain because, beginning with
an irreversible growth simulation, the {\it extra} computational cost
to include detachment is very small for LVST compared to KMC. This is
so because LVST precisely suppresses the time-consuming
detachment/attachment fluctuations that occupy a KMC simulation.
Moreover, the LVST has essentially no restriction on the number of
detachment events simulated during a specific time step.
Quantitatively, Fig.~\ref{scaling} shows that the LVST method requires
only negligibly more run time to include detachment whereas the KMC
simulation cost increases sharply as the detachment rate
increases. The LVST results depend very weakly on the rate of
detachment because the increased cost is associated wholly with
reductions in the step advance rate. Specifically, the adatom density
rises with increasing detachment rate so the gradients evaluated in
Eq.~(\ref{v_grow}) become larger. But the step advance rate is limited
by the condition that the boundary of the level set function can
advance only one grid point in each simulation step.  This implies
that the scaling should be even better for simulations of, say,
annealing processes, where there is no deposition flux and the adatom
density is very low.

\section{Conclusion}

In summary, we have developed a method to model epitaxial growth
including atomic detachment from island edges within the context of
the level set method.  By all reasonable measures, the results are in
excellent agreement with KMC simulations. Moreover, the LVST
simulations scale significantly in CPU-time demand than KMC
simulations when the effective detachment rate is large. This is so
because our mean field method eliminates the many atomic detachment
events (each processed separately in KMC) that do not lead to
successful escape from an island.

\section{Acknowledgments}

We thank M. Kang, C. Andersen, S. Osher, and D. Vvedensky for helpful
discussions. M.P. and A.Z. gratefully acknowledge the support of NSF
DMR-953-1115. This work was also supported by DARPA through
cooperative agreement DMS-9615854 as part of the Virtual Integrated
Prototyping Initiative.

%
% Fig 6
%

%
% Fig 7
%

%
% Fig 8
%

\begin{appendix}

\section{Boundary Conditions}\label{d_vrev}

\begin{figure}
\begin{center}
\psfig{figure=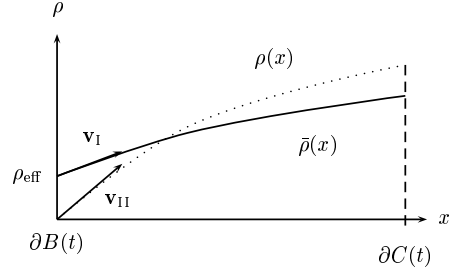,width=7cm}
\end{center}
\caption{\label{a_equi}
   Schematic illustration to the derivation of the equivalence of
   $v^{\rm det}$ and $\rho_{\rm eq}$.
}
\end{figure}

The usual boundary condition for reversible growth \cite{BaZa97} sets
$\rho=\rho_{\rm eq}$ at every island perimeter.  In this Appendix, we
relate $\rho_{\rm eq}$ to the detachment velocity $v^{\rm det}$ used
in this paper.  The idea is to consider two adatom densities:
$\bar{\rho}$ and $\rho$ as schematically shown in
Fig.~\ref{a_equi}. $\bar{\rho}$ ($\rho$) is the system where the
adatom density drops to $\rho_{\rm eq}$ ($\rho$=0) at the island
boundary. We then write down diffusion for both adatom densities and
derive from those the respective reversible growth velocities.

Let $\bar{\rho}$ (the solid line in Fig.~\ref{a_equi}) be the exact
solution of
\begin{eqnarray}
0 & = & D \nabla^2 \bar{\rho}({\rm\bf x})  + F_0 
\quad {\rm\bf x} \in \Omega \label{diff1} \\
\bar{\rho}({\rm\bf x}) & = & \rho_{\rm eq}  
\qquad {\rm\bf x} \in  \partial B(t)
\end{eqnarray}
where $\Omega$ is the domain and $\partial B(t)$ the island
boundaries. $\rho_{\rm eq}$ is the (finite) equilibrium value of
$\bar{\rho}({\rm\bf x})$ at the island boundaries due to detachment
of atoms.

If $C(t)$ is the capture zone and $\partial C(t)$ its boundary (dashed
line in Fig.~\ref{a_equi}), then with the boundary condition
\begin{eqnarray}
\frac{\partial \bar{\rho}}{\partial n} = 0 
\qquad {\bf x} \in \partial C(t) \quad .
\end{eqnarray}
we can uniquely solve the diffusion equation (\ref{diff1}).

In this case the reversible growth velocity is given by
Eq.~(\ref{v_grow}) (labeled ${\rm\bf v}_{\rm I}$ in
Fig.~\ref{a_equi}).

Now we want to replace $\bar{\rho}$ by a adatom density $\rho$ (dashed
line in Fig.~\ref{a_equi}) such that
\begin{eqnarray}
0 & = & D \nabla^2 \rho({\rm\bf x})  + F_0 + F_{\rm rev} 
\quad {\rm\bf x} \in \Omega \label{diff2}  \\
\rho({\rm\bf x}) & = & 0
\qquad {\rm\bf x} \in  \partial B(t)
\end{eqnarray}
were $F_{\rm rev}$ is as explained in Eq.~(\ref{flux_rev}) the
diffusive flux away from the island boundary due to detachment.

It follows from (\ref{diff1}) and (\ref{diff2}) that
\begin{eqnarray}
\nabla^2 \rho({\rm\bf x}) & = & \frac{F_0+F_{\rm rev}}{F_0} 
\nabla^2 \bar{\rho}({\rm\bf x}) 
\end{eqnarray}
and therefore
\begin{eqnarray}
\rho({\rm\bf x}) & = & \frac{F_0 + F_{\rm rev}}{F_0} 
(\bar{\rho}({\rm\bf x}) - \rho_{\rm eq}) 
\label{res1} \quad .
\end{eqnarray}
If we want the two systems $\bar{\rho}$ and $\rho$ to be equivalent,
then we must require 
\begin{eqnarray}
\int_{C(t)} \rho({\rm\bf x}) \,d^2 x & = & 
\int_{C(t)} \bar{\rho}({\rm\bf x}) \,d^2 x \label{int1}
\end{eqnarray}
Combining (\ref{res1}) and (\ref{int1}) we obtain
\begin{eqnarray}
\frac{F_0 +F_{\rm rev}}{F_0} \int_{C(t)} 
\left[\bar{\rho}({\rm\bf x}) - \rho_{\rm eq} \right] d^2x
& = & 
\int_{C(t)} \bar{\rho}({\rm\bf x}) d^2 x 
\end{eqnarray}
which is equivalent to
\begin{eqnarray}
F_{\rm rev} \int_{C(t)} \bar{\rho}({\rm\bf x}) d^2 x & = & 
(F_0+F_{\rm rev}) \int_{C(t)} \rho_{\rm eq}  d^2 x \label{int2} \quad .
\end{eqnarray}
If $A(C)$ is the area of the capture zone, the relation between
$F_{\rm rev}$ and $\rho_{\rm eq}$ is
\begin{eqnarray}
F_{\rm rev} & = & \frac{F_0}{\frac{1}{A(C)}\int_{C(t)} 
\frac{\bar{\rho}({\rm\bf x})}{\rho_{\rm eq}} d^2x -1} \quad .
\end{eqnarray}
Assuming that $R_{\rm det}$ is independent of $d\Gamma$ we can rewrite
Eq.~(\ref{loss_total}) as
\begin{eqnarray}
F_{\rm rev} & = & \frac{1}{\Omega} R_{\rm det} L \quad .
\end{eqnarray}
Then, Eq.~(\ref{vs_def}) reduces to
%
%Together with Eq.~(\ref{loss_total}) and Eq.~(\ref{vs_def}) we can
%write down the relation between $\rho_{\rm eq}$ and $v^{\rm det}$ 
%
\begin{eqnarray}
v^{\rm det}
& = &  a^2 \frac{\Omega}{L} F_{\rm rev} \nonumber \\ & = & 
a^2 \frac{\Omega}{L} \frac{F_0}{\frac{1}{A(C)}\int_{C(t)}
\frac{\bar{\rho}({\rm\bf x})}{\rho_{\rm eq}} d^2x -1}
\quad .
\end{eqnarray}

\section{Derivation of $p_{\rm esc}$}\label{d_pesc}

Any atom that detaches from an island but does not reach the capture
zone boundary is driven by the diffusion field back to the island. In
such a case, $p_{\rm esc}$ is zero. But if the detached adatom does
reach the border of the capture zone, $p_{\rm esc}$ is one.
Therefore, to determine the probability distribution $P({\bf x})$ for
all possible paths of an detaching adatom, we solve the diffusion
equation
\begin{eqnarray}
\nabla^2 P({\rm\bf x}) & = & 0 
\label{a_d1}
\end{eqnarray}
with the boundary condition
\begin{eqnarray}
P({\rm\bf x}) & = & \left\{
\begin{array}{l@{\quad : \quad}l}
0 & {\rm\bf x} \in S_{\rm is} \\
1 & {\rm\bf x} \in S_{\rm cz}
\end{array}\right.
\label{a_d2}
\end{eqnarray}
where $S_{\rm is}$ and $S_{\rm cz}$ are the border of the island and
of the capture zone respectively. These boundary conditions represent
the either successfully escape (probability 1 to reach $S_{\rm cz}$) or
the reattachment to the island (0 at $S_{\rm is}$).

Eq.~(\ref{a_d1}) and (\ref{a_d2}) can in principle be solved for any
island and capture zone geometry. But in our calculations we assume
for simplicity a spherical average for both, the islands and the
capture zones. For this case the general solution to
Eq.~(\ref{a_d1},\ref{a_d2}) is
\begin{eqnarray}
P({\rm\bf x}) = \frac{\log(|{\rm\bf x}|/R_{\rm is})}
{\log(R_{\rm cz}/R_{\rm is})}
\end{eqnarray}
where $R_{\rm is}$ and $R_{\rm cz}$ are the radii of the island and of
the capture zone. From this we obtain the escape probability of a
detached atom by taking $|{\rm\bf x}| = R_{\rm is} + a$ ($a$ is the
lattice parameter) because when an adatom detaches, it is roughly at
distance $a$ from the island boundary.  Therefore
\begin{eqnarray}
p_{\rm esc} & = & \frac{\log((R_{\rm is} + 
a)/R_{\rm is})}{\log(R_{\rm cz}/R_{\rm is})} \quad .
\end{eqnarray}

\end{appendix}

%
% Fig 9
%


\begin{thebibliography}{99}
\bibitem{caveat} This is not strictly true since such detachment/re-attachment processes
can lead to overall diffusion of a two-dimensional island. We ignore this possibility in 
the present work.
\bibitem{comment-adatom}
Note that this gives a lower bound to $\lambda$, since Eq.~(\ref{kink}) is
  derived assuming an infinite step and therefore corner atoms are explicitly
  excluded. Results in Ref.~\onlinecite{CaE99} are checked against KMC steady
  state.
\end{thebibliography}
\end{document}